# Personalization of Code Readability Evaluation Based on LLM Using Collaborative Filtering

Buntaro Hiraki
*Kindai University*
Higashi-osaka, Japan
2333340446r@kindai.ac.jp

Kensei Hamamoto
*Kindai University*
Higashi-osaka, Japan
2433340478s@kindai.ac.jp

Ami Kimura
*Kindai University*
Higashi-osaka, Japan
2110370105t@kindai.ac.jp

Masateru Tsunoda
*Kindai University*
Higashi-osaka, Japan
tsunoda@info.kindai.ac.jp

Amjed Tahir
*Massey University*
Palmerston North, New Zealand
a.tahir@massey.ac.nz

Kwabena Ebo Bennin
*Wageningen UR*
Wageningen, Netherlands
kwabena.bennin@wur.nl

Akito Monden
*Okayama University*
Okayama, Japan
monden@okayama-u.ac.jp

Keitaro Nakasai
*OMU College of Technology*
Osaka, Japan
nakasai@omu.ac.jp

*Abstract*—Code readability is an important indicator of software maintenance as it can significantly impact maintenance efforts. Recently, LLM (large language models) have been utilized for code readability evaluation. However, developers have different ideas of readability and varying skill levels, and therefore the personalization of LLM evaluation would be needed. This study proposes a method which calibrates the evaluation, using collaborative filtering. In our preliminary analysis, the absolute error of the evaluation improved from 1.05 to 0.82, and that suggests the method effectively enhances the accuracy of the readability evaluation using LLMs.

## I. Introduction

Estimating maintenance effort is crucial for managing software projects [2]. Accurate estimation is needed for valid project planning. Code readability is an important factor in software maintainability (i.e., especially with regard to maintenance effort) [1]. This is because developers must first read and understand the target modules to maintain software modules. Prior research considered readability to evaluate the maintainability of modules which establishes the need to evaluate code readability for maintenance effort estimation [2].

Various metrics have been proposed to evaluate the readability of modules [3]. For example, Mi et al. classified modules based on their readability using a convolutional neural network [5]. Recently, LLMs have been applied to various software development activities to improve accuracy. Simões et al. applied LLM to evaluate the readability of the modules by ordinal scale and compared the performance with conventional approaches [8].

In this study, we focus on the necessity of personalization for the readability evaluations. The readability of modules could be different among software developers [3], This is because developers could have different ideas of readability and varying skill levels. That is, one developer might judge a module as having high readability, whereas it might not be easy for other developers to read it. As a result, the evaluation of the readability by LLMs could not be valid for some developers, even if the evaluation is very accurate.

Our study aims to enhance the validity of LLMs' readability evaluations. To achieve this goal, we propose calibrating the evaluations using mathematical models for each developer and suppressing the differences between LLMs and developers. Our proposed approach personalizes readability evaluations. This is because the approach is a sort of recommender system, and it is regarded as personalization [7].

Even if we use conventional approaches such as software metrics to evaluate readability, personalization is necessary when the evaluation differs among developers. To the best of our knowledge, no study applies personalization to code readability. While using LLM, a study [4] generated personalized source code. Our approach is different and in contrast to the study of [4].

## II. Personalization of Readability

### A. Calibration of Evaluation by LLM

Our approach builds calibration models for readability evaluation by LLM. The dependent variable is the evaluation by developers. Typically, the variable is ordinal scale. The independent variables are the evaluation by the LLMs and metrics of modules such as lines of codes and complexity. When values of the dependent variable are stored, the calibration model can be built because values of the independent variables are stored automatically. Figure 1 shows an overview of the calibration model. In the figure, the readability of modules A, B, C, ... are evaluated by developer α, and the model is calibrated.

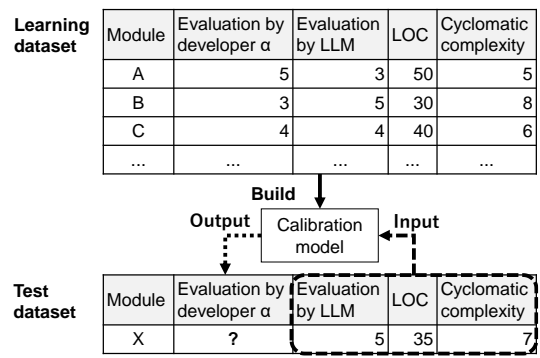

Fig. 1  **Overview of the calibration model**

*B. Personalization Using Collaborative Filtering*

To build an accurate calibration model for each developer, readability evaluation by the developers should be stored to some extent. However, every developer does not evaluate the readability of many modules and store the evaluations. Therefore, it is challenging to build calibration models for such users. The easiest solution to the problem is to utilize the calibration models of other developers. When readability evaluation scores are similar to the other developers, the model's accuracy is expected to be high.

The situation is similar to e-commerce sites. On e-commerce sites, there are many articles, but it is unclear which articles are suitable for each user. To help select and recommend suitable articles, the system based on collaborative filtering [9] recommends articles that are also suitable for the user whose preference is similar to the target user. Based on the concept of the recommender system, we select a suitable calibration model for each developer, regarding the models as the articles.

Our approach selects the most suitable calibration model for each developer by the following procedure (see Figure 2):

1. Build calibration models based on developers' evaluation, which is stored enough to build an accurate model.
2. A target developer evaluates the code readability of several modules.
3. The similarity of the target developer to the developers used in Step 1 is calculated based on readability evaluation.
4. Select the calibration model that is built based on the most similar developer identified in Step 3.

We assume that the amount of required evaluation in step 2 is less than in step 1. Note that the evaluation by a similar developer is not used directly in most cases. For instance, the target developer can refer to the evaluation of module B in Figure 1 but cannot refer to that of X. Since the readability of module X is not evaluated by both the similar and target developer.

## III. EXPERIMENT

We used dataset by Scalabrino [6] which includes the readability evaluation of 200 code snippets with five grades by nine participants. We used ChatGPT3.5 as the selected LLM, and input the prompt "Please rate readability of the following source code by five-grades" to evaluate 200 snippets. On the dataset, the evaluation was different among developers (Standard deviation of the evaluation on each snippet was about 0.9). This suggests there is no universal evaluation of readability. This is why we did not performed prompt engineering to enhance accuracy of LLM. Although we used code snippets instead of modules, we could utilize the approach by aggregating readability evaluation of snippets to approximate the evaluation of a module.

We applied the leave-one-cross validation method in the experiment. As evaluation criteria of the accuracy of the calibration model, we used absolute value between the calibrated value and the actual value that participants evaluated. The procedure of the experiment is described as follows:

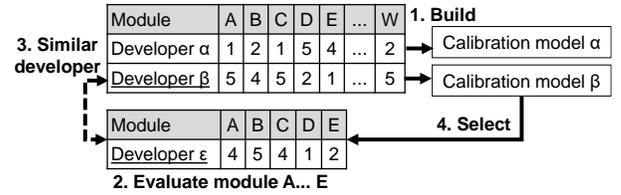

Fig. 2 **Procedure of our proosed approach**

TABLE I. ABSOLUTE ERROR OF EACH APPROACH

| LLM | Random selection | Cosine similarity | Euclidean distance |
|---|---|---|---|
| 1.05 | 0.92 | 1.09 | 0.82 |

i. Split 200 snippets into 140 and 60 ones. The former is used as a learning dataset, and the latter is a test dataset.
ii. Calculate the similarity of participants based on the readability of 30 snippets, which includes the learning dataset.
iii. Select the calibration model of the other participants which is most similar to the target participants.
iv. Using the selected model, the readability of 60 snippets included in the test dataset is predicted.

To compute similarity, we used cosine similarity and Euclidean distance. As the baseline, we selected calibration models randomly for each participant.

The result is shown in Table 1. Euclidean distance was used, and absolute error was the smallest of the methods. The result shows that the personalization of readability evaluation is needed, and the calibration model is effective compared to using LLM directly. When the calibration model was randomly selected, the absolute error was smaller than LLM. This might suggest that calibration of LLM is needed, regardless of personalization. The accuracy was the lowest when we used cosine similarity.